\documentstyle[11pt,newpasp,twoside,psfig]{article}
\begin{document}

\title{AGN as cosmic thermostats}
\author{Christian R. Kaiser}
\affil{University of Southampton, Southampton, S17 1BJ}
\author{Marcus Br{\"u}ggen}
\affil{International University Bremen, Campus Ring 1, 28759 Bremen}
\author{James J. Binney}
\affil{University of Oxford, 1 Keble Road, Oxford OX1 3NP}

\setcounter{page}{111}
\index{Kaiser, C.R.}
\index{Br{\"u}ggen, M.}
\index{Binney, J.J.}

\begin{abstract}
We present a simple analytical model and the results of numerical
simulations supporting the idea of periodical heating of cluster gas
by AGN outflows. We show why, under this assumption, we are extremely
unlikely to observe clusters containing gas with a temperature below
about 1\,keV. We review the results from numerical hydro-simulations
studying the mechanisms by which AGN outflows heat the cluster gas.
\end{abstract}

\section{How to find clusters with gas below 1\,keV}

We assume a NFW profile for the gravitational potential created by the
dark matter of a galaxy cluster. Adopting a simple power-law
prescription for gas mass as a function of entropy for the cluster
gas, we can solve the hydrodynamical equations describing hydrostatic
equilibrium. We then let the cluster gas cool radiatively during a
small timestep. This results in a slightly modified mass-entropy
distribution of the gas. For this new distribution we again find the
hydrostatic solution and repeat the whole process until the gas at the
centre of the cluster has radiated away all its energy. This happens
after about 300\,Myr. At this point we expect the cold gas to trigger
a new outburst of the AGN which will heat the gas back to a state
resembling the initial configuration. It is interesting to note that
during the evolution of the cluster gas the mass-entropy distribution
is always very closely approximated by a power-law. This is even true
in the case of an initially flat distribution, i.e. all gas in the
cluster has the same entropy. Using this simple model, we calculate
the fraction of clusters in a survey which contains an amount of gas
with temperature below 1\,keV detectable in an X-ray observation ten
times deeper than the original survey observations. The result is that
only one in a few hundred clusters should be detected to contain such
cold gas. If the idea of AGN heating phases separated by long cooling
phases is the correct picture for the evolution of the cluster gas,
then it is not surprising that we do not observe any cold gas in
clusters. Details of the model with a more detailed discussion can be
found in Kaiser \& Binney (2002).

\section{How to heat a cluster with an AGN}

\begin{figure}[t]
\centerline{
\psfig{figure=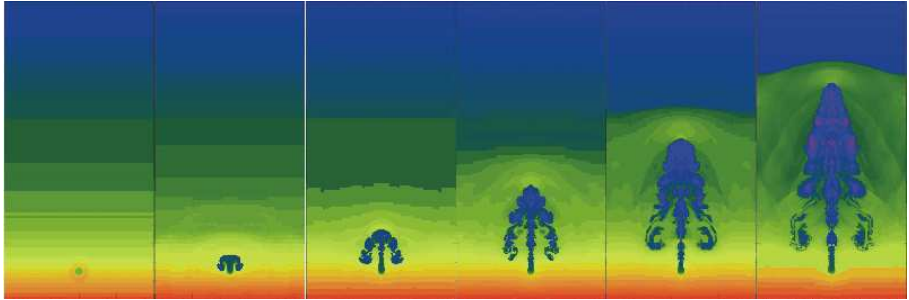,width=12cm,angle=-90}
}
\caption{Snapshots of a numerical simulation showing the interaction
of an AGN outflow with a stratified cluster atmosphere. Thermal energy
is injected in the spherical volume visible close to the bottom of the
computational domain. The colour indicate gas density on a logarithmic
scale. Individual snapshots are separated by about 50\,Myr.}
\end{figure}

The idea of repeated heating phases immediately leads to the question
of how AGN outflows heat the cluster gas. The problem is intrinsically
complex as fluid instabilities will lead to turbulent mixing of the
outflow material with the cluster gas. Because of the intrinsic
3-dimensionalityof the problem, analytical studies are virtually
impossible. Therefore we have performed hydrosimulations to study the
heating of the cluster gas by AGN outlfows. Figure 1 shows a
timeseries from one of our simulations. We were able to surpress
purely numerical effects like numerical diffusion which have plagued
earlier simulations through the use of adaptive mesh refinement. We
were able to show that even the uplift of cold gas from the cluster
centre and large-scale mixing of outflow material and cluster gas can
significantly prolong the radiative cooling time of the cluster
gas. Further simulations including the creation of entropy through
small-scale mixing are required for more realitistic
scenarios. Details of the simulations and a discussion of the results
can be found in Br{\"u}ggen \& Kaiser (2002).

\section{Conclusions} 

We conclude that the scenario of cluster gas periodically heated by
AGN outflows can explain the observed lack of cold cluster gas. We
also find that outflows from AGN can indeed heat the cluster gas
sufficiently to counteract radiative cooling.


\begin{references}
\reference Br{\"u}ggen, M. and Kaiser, C.R.: 2002, Nature, 418, 301
\reference Kaiser, C.R. and Binney, J.J.: 2002, MNRAS: accepted,
astro-ph/0207111 

\end{references}
\end{document}